\documentclass[preprint,12pt]{elsarticle}
\usepackage{graphicx}
\usepackage{epstopdf, epsfig,mathtools,bigints}
\usepackage{amsmath}

\usepackage{lineno}
\modulolinenumbers[1]

\begin{document}

\begin{frontmatter}

\title{Analytical models for the asymmetric wake of vertical axis wind turbines}

\author[1,2]{Pablo Ouro\corref{cor1}} \ead{pablo.ouro@manchester.ac.uk}
\address[1]{School of Mechanical, Aerospace and Civil Engineering, University of Manchester, Manchester, M13 9PL, UK}
\address[2]{Hydro-environmental Research Centre, School of Engineering, Cardiff University, Cardiff, CF24 3AA, UK}
\cortext[cor1]{Corresponding author: Dr Pablo Ouro (pablo.ouro@manchester.ac.uk)}
\author[3]{Maxime Lazennec} \ead{maxime.lazennec@polytechnique.edu}
\address[3]{\'Ecole Polytechnique, 91120 Palaiseau, Paris, France}

\begin{abstract}
Arrays of Vertical Axis Wind Turbines (VAWTs) can achieve larger power generation per land area than horizontal axis turbines farms, due to the positive synergy between VATs in close proximity.
Theoretical wake models enable the reliable design of the array layout that maximises the energy output, which need to depict the driving wake dynamics. 
VAWTs generate a highly complex wake that evolves according to two governing length-scales, namely the turbine rotor's diameter and height which define a rectangular shape of the wake cross-section, and feature distinct wake expansion rates.
This paper presents analytical VAWT wake models that account for an asymmetric distribution of such wake expansion adopting a top-hat and Gaussian velocity deficit distribution. 
Our proposed analytical Gaussian model leads to an enhanced initial wake expansion prediction with the wake width ($\varepsilon$) behind the rotor equal to $(\beta/4 \pi)^{1/2}$ with $\beta$ being the ratio of initial wake area to the VAWT's frontal area, which addresses the limitations of previous models that under-predicted the wake onset area.
Velocity deficit predictions are calculated in a series of numerical benchmarks consisting of a single and an array of four in-line vertical axis wind turbines.
In comparisons with field data and large-eddy simulations, our models provide a good accuracy to represent the mean wake distribution, maximum velocity deficit, and momentum thickness, with the Gaussian model attaining the best predictions.
These models will aid to drive the design of VAT arrays and accelerate this technology.
\end{abstract}

\begin{keyword}
Vertical Axis Wind Turbines 
\sep Wakes 
\sep Self-similarity 
\sep Large-Eddy Simulation 
\sep Wind farm
\sep VAWT
\end{keyword} 

\end{frontmatter}


\section{Introduction}


In the global landscape of wind energy generation, all large-scale wind farm projects comprise Horizontal Axis Wind Turbines (HAWTs) as a well-established technology for both onshore and offshore environments, becoming one of the most cost-effective resources to harness renewable energy \citep{Veers2019}. 
Conversely, Vertical Axis Wind Turbines (VAWTs) are being developed at a much slower pace with the remaining main challenge to prove their financial viability, conditioned by the need for enhancing their power generation capabilities.
VAWTs offer a series of advantages over their HAWT counterparts that can lead to innovative wind and hydro-kinetic energy projects, unfeasible if HAWTs were the chosen technology. 
For instance, they can effectively harness kinetic energy from relatively low-to-medium flow velocity ranges as those found in urban areas, rivers or tides, or be adopted in environmentally sensitive regions as their slower rotational speeds can reduce fish collision risk or acoustic contamination.


VAWT rotor blades rotate in a plane perpendicular to the approaching flow direction, generating vortex structures over the upwind rotation half and interacting with these over the downwind one. 
Despite such fluid-structure complexity of VAT wakes, the research in characterising VAWT wakes is still insufficient with limited knowledge about the governing flow mechanisms in the far-wake dynamics. 
This led to extrapolate features of HAT wakes into the dynamics of VAWT wakes which are fundamentally unequal, e.g. tip or dynamic-stall vortices patterns \citep{Liu2019}.

Most experimental tests of VATs look at how to improve their performance, leading to a limited number of studies analysing the near-wake dynamics and an almost absence of extensive tests that investigate the far-wake.
Few examples of the latter at small-scale laboratory scale are: 
\citet{Rolin2018} measured with Particle Image Velocimetry (PIV) up to 10 diameters ($D_0$) downstream of a VAT analysing the mean and turbulent kinetic energy equations; 
\citet{Araya2017} investigated the near- to far-wake transition for turbine rotors with different number of blades and tip-speed ratios using PIV to measure up to 11$D_0$ downstream; and 
\citet{Ouro2019RE} measured the wake up to 14$D_0$ downstream with acoustic Doppler velocimeter showing that remnants of the turbine-induced wake are still observed at such far distances downstream.

\begin{figure}
\centerline{\includegraphics[width=\linewidth]{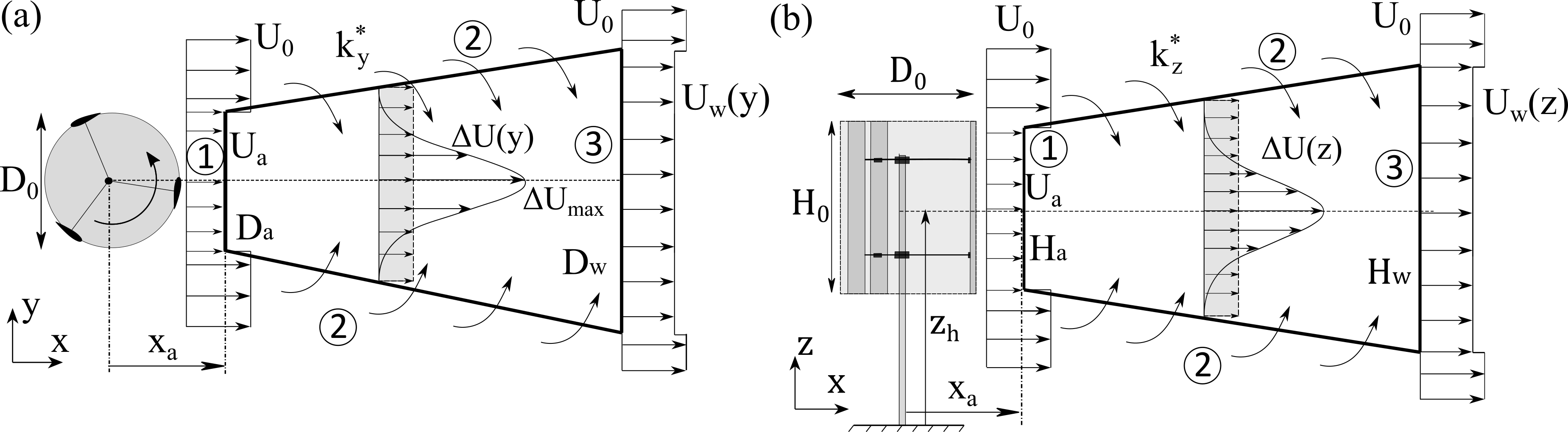}}
\caption{Wake evolution behind a VAWT of diameter $D_0$ and height $H_0$ over the (\textit{a}) horizontal ($xy$) plane at a mid-height elevation $z_h$ from the ground surface, and (\textit{b}) vertical ($xz$) plane through the rotor's centre, with the thick solid line denoting the control volumes of interest.}  \label{fig:sketch_2}
\end{figure}

The complex fluid dynamics in VAT wakes depend on the operational regime accounted for by the Tip-Speed Ratio (TSR) which relates the blades' angular speed to the free-stream velocity, and the turbine's rotor geometric solidity, which indicates the proportion of the swept perimeter occupied by the blades \citep{Parker2016,Posa2020a}.
VAWTs designed with low solidity rotors operate at high TSR which leads the blades to undergo light dynamic stall, i.e. flow separation occurs for effective angles of attacks larger than the static stall angle but there is no full detachment of the generated leading-edge vortices preventing a sudden drop in lift and torque.  
Alternatively, rotors with a higher solidity operate at low TSR and blades experience deep dynamic stall, meaning the attached leading-edge vortices enhance the lift-generation capabilities of the blades but at a given angle that is normally when torque generation is maximum these energetic vortices are shed \citep{Ouro2017CAF,Posa2020a}.
Whilst the latter configurations are common for Vertical Axis Tidal Turbines (VATTs), the design of VAWTs normally adopt with a low solidity rotor and attain their optimal TSR, i.e. relative rotational speed at which their power coefficient is maximum, in the range of 4--5 \citep{Tescione2014}, but these may suffer from self-starting issues.
Hence, designers are adopting more compact, higher solidity rotors that operate at optimal TSR ranges of approx. 1.5--2.5, which is mostly adopted in VATTs for hydro-kinetic applications at sea or in rivers \citep{Brochier1986,Ouro2017CAF,Bachant2016}. 

The horizontal-to-vertical asymmetric distribution of VAWT wakes is depicted in Fig. \ref{fig:sketch_2}, which considers a control volume behind a turbine with diameter $D_0$ and height $H_0$ over which mass and momentum needs to be conserved \citep{Bastankhah2014}. 
Over the horizontal $xy$-plane, the main contribution to momentum entrainment over the lateral boundaries of the control volume results from the blade-induced energetic vortices \citep{Kadum2020}.
This differs from the wake recovery dynamics over the $xz$-vertical plane (Fig. \ref{fig:sketch_2}b), as shear layers generated in this direction result from the tip vortices \citep{Tescione2014}.
These blade-induced structures are not identical leading to potentially dissimilar wake mixing rates over the horizontal ($k_y^*$) and vertical ($k_z^*$) planes, causing an asymmetric wake recovery.
Furthermore, VAWT rotors with height-to-diameter aspect ratio different to unity leads to having two length-scales with which the wake dynamics, e.g. velocity distributions, can be characterised, unlike HAWT wakes that scale with its diameter \citep{Shamsoddin2020}.
As shown in Fig. \ref{fig:sketch_2}, in the horizontal plane $D_0$ is the characteristic length scale while in the vertical plane this is $H_0$.


More insights into the complex turbulence structures generated by vertical axis wind or tidal rotor blades were gained from numerical simulations. 
Research has evidenced the need for eddy-resolving closures, such as Large-Eddy Simulation (LES), to resolve the flow within the VAT rotor and its wake, as Reynolds-Averaged Navier Stokes models can fail to predict dynamic stall \citep{Ouro2018JFE} or the replenishment of mean kinetic energy in the wake which is driven by turbulence fluctuations \citep{Bachant2016,Ouro2019RE,Posa2020b}. 
Such flow complexity requires fine numerical grids with three approaches mostly adopted to represent VAT rotors.
Geometry-resolved simulations, e.g. adopting an immersed boundary method, provide a high-resolution at the rotor capturing the vortices induced by the blades \citep{Posa2018,Ouro2017CAF}.
Actuator techniques enable to adopt lower grid resolution while capturing the main wake dynamics but fail at representing the dynamic stall vortices 
\citep{Porte-Agel2020}.
Actuator Line Method (ALM) is the most widely adopted \citep{Abkar2017,Shamsoddin2014,Shamsoddin2016} whilst the Actuator Surface Method (ASM) can provide further gains to improve the resolution of the flow at rotor level \citep{Massie2019}.

There is still a need for developing analytical wake models tailored to VATs that enable its prediction at almost no computational cost \citep{Meneveau2019}, which is the aim of this article. Analytical models are widely adopted to design array layouts by industry and researchers as they enable the evaluation of multiple operational scenarios within a reasonable time frame \citep{Stansby2016}, thus the presented models will be invaluable for the future design of VAT arrays \citep{Stevens2017}.
Our models improve the wake expansion both in terms of expansion rate and wake onset area, and account for an asymmetric wake recovery, relevant for VATs with height-to-diameter aspect ratios different to unity.
The structure of the paper is: \S\ref{sec:problem} describes the underpinning basis our wake models are built upon, which are derived in \S\ref{sec:models}.
Validation in four cases with comparisons to large-eddy simulation results, field data and other available models are presented in \S\ref{sec:results}, with the main conclusions drawn in \S\ref{sec:conclusions}.



\section{Novelty of wake models}  \label{sec:problem}

The underpinning physics of our models for VAWT wakes lie in considering the wake to expand according to two length-scales, namely the wake diameter ($D_w$) and height ($H_w$), featuring a rectangular shape analogously to the projected area of the VAWT ($A_0=D_0H_0$), as shown in Fig. \ref{fig:sketch_2}.
The cross-sectional wake area $A_w=D_wH_w$ at any downstream location ($x$) is determined as the evolution of the wake from its onset location, $x_a$, corresponding to the location at which pressure equilibrium is reached, i.e. the pressure at the wake centre is the same as the free-stream one.
The latter needs to be downstream of the swept perimeter region, i.e. $x/D_0 \geq 0.5$, and for simplicity we set $x_a=0.5D_0$ as the location of the wake onset in our models, analogously to the hypothesis done in HAT wake models in which such pressure equilibrium is attained almost immediately downstream of the turbine rotor \citet{Frandsen2006}.

The wake expansion is determined from two non-dimensional parameters corresponding to the horizontal ($k_y$) and vertical ($k_z$) directions, whose values can be dissimilar as the shear layers generated by tip vortices over the vertical direction are different to those from dynamic-stall vortices on the horizontal one. 
Our wake expansion considerations allow to develop a Gaussian wake model that accounts for a physically valid estimation of the initial wake area, addressing limitations found in other VAT models partly build upon concepts from HAWT wakes whose under-prediction of the wake area at $x_a$ leads to non-physical values of velocity deficit in the near-wake, as discussed in \S\ref{sec:results}.
Note that from hereafter we will use VAT to avoid distinction of wind or tidal turbines as models can be applicable to either of these.



\section{Derivation of the wake models} \label{sec:models}

\subsection{Momentum conservation}

We build our momentum conserving wake models starting from the conservation form of the Reynolds Averaged Navier-Stokes equation for high Reynolds numbers in the streamwise direction. After neglecting the pressure and viscous terms, this equation reads

\begin{equation}
\frac{\partial U_w (U_0 - U_w)}{\partial x} + \frac{\partial V_w (U_0 - U_w)}{\partial y} + \frac{\partial W_w (U_0 - U_w)}{\partial z} =
\frac{\partial u'u'}{\partial x} +\frac{\partial u'v'}{\partial y} + \frac{\partial u'w'}{\partial z}      \label{eq:RANS}
\end{equation}

with ($U_w,V_w,W_w$) being the vector of mean wake velocities in the streamwise, transverse and vertical directions respectively, and $u'u'$, $u'v'$ and $u'w'$ denote time-averaged turbulent fluctuation correlations.
We integrate \ref{eq:RANS} at any streamwise location of a control volume that embeds the turbine and expands
over the $y$ and $z$ directions from $-\infty$ to $\infty$, which together with the assumption the shear stresses vanish when increasing the distance from the wake centre provides the resulting RANS equation,

\begin{equation}
  \frac{d}{dx} 
  \bigintsss_{-\infty}^{\infty} \left(U_{w}\left(U_{0} - U_{w}\right) -u'u'\right)\mathrm{d} A \approx 0       \label{eq:RANSint}
\end{equation}

The streamwise variation of $u'u'$ is much reduced when compared to the convective term \citep{Bastankhah2014}, hence \ref{eq:RANSint} can be simplified to obtain the momentum integral \citep{Tennekes} as

\begin{equation}
    \rho\bigintssss_{-\infty}^{\infty}  \left(U_{w}\left(U_{0} - U_{w}\right)\right)\mathrm{d} A = T       \label{eq:Thrust1}
\end{equation}

This equilibrium condition states that the momentum deficit flux in the wake is proportional to the thrust force $T$ exerted by the turbine.
Note that we consider the wakes to be in zero-pressure gradient flow and the incoming velocity to be mostly uniformly distributed in the three directions of space.
The thrust force can be related to the thrust coefficient ($C_T$) from the actuator disk theory as

\begin{equation}
    T = \frac{1}{2}C_{T}\rho A_{0} U_{0}^2      \label{Thrust2}
\end{equation}

\subsection{Top-hat wake model}


The value of the wake onset area $A_{a}$ can be determined using the actuator disk theory as $A_{a} = \beta A_{0}$, with $\beta$ representing the relative initial wake expansion at $x_a$ in terms of the turbine rotor's cross-section $A_{0}$, independently of whether such cross-section is circular (HATs) or rectangular (VATs).
The actuator disk theory states the velocity over the plane at $x_{a}$ is $U_{0}(1-2a)$ whilst at the rotor centre plane the velocity is $U_{0}(1-a)$, with $a=\frac{1}{2}\left(1-\sqrt{1-C_{T}}\right)$ being the so-called induction factor. 
Hence, the value of $\beta$ is determined based on energy conservation as:

\begin{equation}
    \beta = \frac{A_{a}}{A_{0}}=\frac{1-a}{1-2a}=\frac{1}{2}\frac{1+\sqrt{1-C_{T}}}{\sqrt{1-C_{T}}}
    \label{eq:beta}
\end{equation}

The wake area at any streamwise location is determined similarly to the approach presented by \citet{Frandsen2006} for HATs with $D_w \propto x^{1/2}$.
Hence, the wake width is considered to expand asymmetrically over the horizontal and vertical directions as:

\begin{equation}
    D_{w} = D_0 \left(\beta + k_{w_{y}}\frac{x-x_a}{D_0}\right)^{1/2}  \quad\text{, }\quad 
    H_{w} = H_0 \left(\beta + k_{w_{z}}\frac{x-x_a}{H_0}\right)^{1/2}  \label{eq:FranHw}
\end{equation}


We now apply momentum balance to a control volume that embeds the operating turbine expanding some distance upstream and downstream 
to obtain the velocity deficit $\Delta U = U_0-U_w$ for the top-hat model with asymmetric expansion over the horizontal and vertical directions:


\begin{equation}
 \frac{\Delta{U}}{U_{0}} 
 = \frac{1}{2}\left(1-\sqrt{1-\frac{2 C_{T}}{
 \left(\beta + k_{w_{y}}\frac{x-x_a}{D_0}\right)^{1/2} 
 \left(\beta + k_{w_{z}}\frac{x-x_a}{H_0}\right)^{1/2}
 }}\right) \label{eq:FranVAT}
\end{equation}

Values of $k_{w_y}$ and $k_{w_z}$ are considered equal to 2.0$I_u$, with $I_u$ denoting streamwise turbulence intensity. 
This value is half the one adopted in HAT wakes \citep{Frandsen2006} and in other VAT wake models \citep{Abkar2019}.
We propose this value as it provides better predictions of momentum thickness or maximum $\Delta U$ as shown later in \S\ref{sec:results}.

\subsection{Gaussian wake model}

A cornerstone in the derivation of a Gaussian wake model is to assume the velocity deficit distribution to be self-similar, i.e. at any streamwise distance $\Delta U$ can be directly determined from local scales of velocity and length \citep{Tennekes}.
For axisymmetric wakes, as is the case of HAT wakes \citep{Stallard2015}, self-similarity is attained if the transverse distribution of $\Delta U$ consistently follows a given function $f(-\frac{1}{2}(r/\sigma)^2$), where $r/\sigma$ is the distance from the wake centre ($r$) normalised by the characteristic wake width ($\sigma$); and a given velocity scale $C(x)$, which is determined as the maximum normalised velocity deficit ($\Delta U_{max}/U_0$) at any streamwise distance.
The self-similar normalised $\Delta U$ can be written as 

\begin{equation}
    \frac{\Delta U}{U_0} = \frac{U_0 - U_w}{U_{0}} = C(x) f \left(-\frac{1}{2}\frac{r^2}{\sigma^2} \right)   \label{eq:selfsimi}
\end{equation}



We assume $\Delta U$ to be self-similar following a Gaussian shape function, with results in \S\ref{sec:results} showing this condition is deemed valid.
Adopting a Gaussian distribution allows a more physically-realistic description of the wake velocity deficit compared to top-hat models that assume a uniform value across the wake width.
Whilst in most HAT wake cases the self-similar shape function is defined only by the turbine diameter \citep{Bastankhah2016,Shapiro2018},
for VATs both its diameter and height are characteristic length-scales that determine the distribution of $\Delta U/U_0$ over the horizontal and vertical directions of the wake.
The wake asymmetry is accounted in our model by adopting the superposition of two Gaussian distributions $\Delta U(y)$ and $\Delta U(z)$, whose characteristic wake widths $\sigma_y$ and $\sigma_z$ scale depending on $D_0$ and $H_0$, respectively.
Following \citet{Bastankhah2014}, we propose these wake widths to expand linearly in the downwind direction:

\begin{eqnarray}
\frac{\sigma_{y}}{D_0} = k_y^{*} \frac{x-x_a}{D_0} + \varepsilon_{y}    \quad\text{, }\quad 
\frac{\sigma_{z}}{H_0} = k_z^{*} \frac{x-x_a}{H_0} + \varepsilon_{z}     \label{sigmaz}
\end{eqnarray}

Wake expansion rates $k_y^*$ and $k_z^*$ are estimated as 0.35$I_u$ as proposed by \citet{Bastankhah2014}, which are proven adequate given the wake predictions presented in \S\ref{sec:results} considering $\varepsilon_y$ and $\varepsilon_z$ that represent the wake onset width are well determined.
The wake velocity \ref{eq:selfsimi} can be re-written as

\begin{equation}
 U_{w}=U_{0}\left(1-C(x) \text{exp}{\left(-\frac{y^2}{2 \sigma_{y}^2}-\frac{z^2}{2 \sigma_{z}^2}\right)}\right) \label{eq:GaussVATUw}
\end{equation}

This is an algebraic equation with one unknown, the velocity scale $C(x)$, which is solved equating the momentum integral \ref{eq:Thrust1} to the turbine thrust force \ref{Thrust2} to obtain the definition of the wake velocity \ref{eq:GaussVATUw} as

\begin{eqnarray}
    && \bigintsss_{-\infty}^{\infty}
    \hspace{-0.3cm}
    U_{0}^2C(x) 
    \text{exp}{\left(-\frac{y^2}{2 \sigma_{y}^2}-\frac{z^2}{2\sigma_{z}^2}\right)} 
    \left(1-C(x) 
    \text{exp}{\left(-\frac{y^2}{2 \sigma_{y}^2}-\frac{z^2}{2\sigma_{z}^2}\right)}\right)
    \mathrm{d} A
   \nonumber\\    &&
    =\frac{1}{2}C_{T} A_{0} U_{0}^2 \label{eq:Gaus1}
\end{eqnarray}

Considering the rectangular cross-section of the wakes $A_w$=$D_w H_w$ and 
$\int_{-\infty}^{\infty} \text{exp}(-y^2/(2 \sigma_y^2))$ dy= $\sqrt{2\pi} \sigma_y$ and 
$\int_{-\infty}^{\infty} \text{exp}(-y^2/\sigma_y^2)$dy=$\sqrt{\pi} \sigma_y$, 
\ref{eq:Gaus1} can be integrated to determine the normalised maximum velocity deficit, $C(x)$, as

\begin{equation}
    \sigma_{y} \sigma_{z} \pi C(x)^2 - 2\pi\sigma_{y}\sigma_{z}C(x) + \frac{1}{2}C_{T}A_{0} = 0
\end{equation}


From the two possible solutions to this quadratic equation, the value that provides a physical solution for the characteristic velocity scale is then

\begin{equation}
    C(x) = \frac{\Delta U_{max}}{U_0}= 1 - \sqrt{1-\frac{C_{T}A_{0}}{2\pi\sigma_{y}\sigma_{z}}} \label{eq:GaussCxVAT}
\end{equation}

Our proposed Gaussian model for VAT wakes is obtained from \ref{eq:GaussVATUw} and \ref{eq:GaussCxVAT} as

\begin{equation}
\frac{\Delta{U}}{U_{0}} = \left(1-\sqrt{1-\frac{C_{T}D_0H_{0}}{2\pi\sigma_{y}\sigma_{z}}}\right)
\text{exp} {\left(-\frac{y^2}{2 \sigma_{y}^2}-\frac{z^2}{2 \sigma_{z}^2}\right)}
\label{eq:GaussDUVAT}
\end{equation}

The values of $\varepsilon_y$ and $\varepsilon_z$ are determined from the mass flow deficit rate immediately behind the turbine's rotor at $x_a$ by equating that predicted by the top-hat model \ref{eq:FranVAT} and to the expression from the Gaussian model \ref{eq:GaussDUVAT}, providing the following relation:

\begin{eqnarray}
    \frac{D_aH_a}{2}\left(1-\sqrt{1-\frac{2C_{T}}    \beta}  \right)
     =
    2\pi D_0 \varepsilon_{y} H_0 \varepsilon_{z}\left(1-\sqrt{1-\frac{C_{T}}{2 \pi \varepsilon_{y} \varepsilon_{z}}}\right)
    \label{eq:MassFRgaussian}
\end{eqnarray}

Hence, $\varepsilon_{y}$ and $\varepsilon_{z}$ expressions for VAT wakes are determined as

\begin{equation}
 \varepsilon_{y} \varepsilon_{z} = \frac{\beta}{4 \pi}  
    \label{eq:betaepsilon}
\end{equation}

Note that the initial wake expansion rates are just a function of $\beta$, i.e. only depends on the thrust coefficient $C_T$. 
The lack of an extensive experimental campaign focused on VAT far-wakes prevents from individually accounting for each of these wake expansion rates.
Hence, for simplicity, we assume the normalised wake onset width is identical in the horizontal and vertical directions, i.e. $\varepsilon_{y} = \varepsilon_{z}$, with their value becoming

\begin{equation}
    \varepsilon_y = \varepsilon_z = {\frac{1}{\sqrt{4 \pi}}}\sqrt{\beta}     \label{eq:varepsVAT}
\end{equation}

It is noteworthy that our formulation of the wake onset width (\ref{eq:varepsVAT}) differs from the definition $\varepsilon = 0.25\sqrt{\beta}$ proposed by \citet{Abkar2017} and \citet{Abkar2019}, which corresponds to the value for HATs.
Our formulation overcomes limitations from these models in relation to non-physical estimations of $\Delta U$ at short distances behind the turbines, which resulted from an incorrect definition of $\varepsilon$ in \ref{eq:GaussDUVAT}. 
Comparing $\varepsilon \approx 0.282\sqrt{\beta}$ (\ref{eq:varepsVAT}) and $0.25\sqrt{\beta}$, the latter is approx. 12\% smaller, thus leading to and underestimation of the wake onset width at $x=x_a$. 


\section{Prediction of the wake models} \label{sec:results}

We present the validation of the wake models in four cases with comparisons to laboratory experiments, field data, and large-eddy simulations.
Results from the Gaussian wake model by \citet{Abkar2019}, whose underlying physics were partially based on HATs, are also included to compare the accuracy and reliability of both model predictions. 
From hereafter, the origin of coordinates are at the turbine centre of rotation and mid-height and for simplicity the turbine diameter and height are represented by $D$ and $H$ instead of $D_0$ and $H_0$.

\subsection{Single VAWT operating in a turbulent boundary layer flow} \label{sec:cases1}

Cases 1a to 1c correspond to a single VAWT operating in a turbulent boundary layer flow representing three scenarios in which the device attains different Tip-Speed Ratios (TSR), thrust coefficient ($C_T$), and aspect ratio \citep{Abkar2019}. 
In cases 1a and 1b, the turbine has a diameter $D$ = 26m with a height $H$ = 24 m whilst in case 1c its height is equal to 48 m, which yields diameter-to-height aspect ratios close to unity for former cases and almost of two for case 1c.
We are interested in comparing cases 1a and 1b as the VAWT operates at TSR of 3.8 and 2.5 respectively, attaining thrust coefficients of 0.65 and 0.34 that lead to different wake dynamics.
In case 1c, the turbine has a $C_T$ equal to 0.64 and operates at TSR = 3.8, as in case 1a, but its aspect ratio of two promotes a larger wake asymmetry when comparing its recovery over the horizontal and vertical directions.
Turbines are equipped with three NACA 0018 blades with a chord length $c$ = 0.75 m, leading to a solidity value $N_b c / \pi D_0 \approx $ 3\%. 
The free-stream velocity at hub-height ($U_0$) is 7.0 ms$^{-1}$ with a turbulent intensity ($I_u$) of 9.1\%.

\begin{figure}
\centerline{\includegraphics[width=0.8\linewidth]{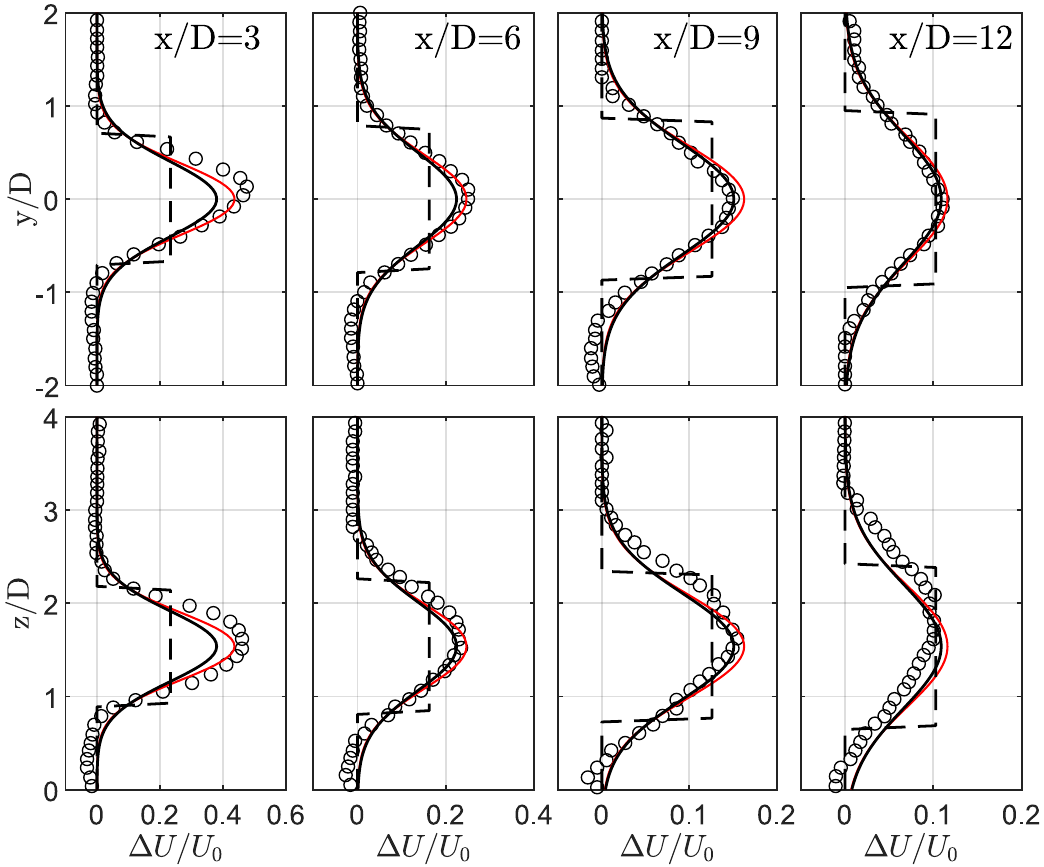}}
\caption{Normalised velocity deficit profiles for case 1a with $C_T$ = 0.65, TSR = 3.8, $D$ = 26 m and $H$ = 24 m. Comparison of our proposed top-hat (dashed line) and Gaussian (solid black line) analytical wake models, with the Gaussian model proposed by \citet{Abkar2019} (solid red line) and LES-ALM results from \citet{Abkar2017} (circles).}  \label{fig:cases1profiles}
\end{figure}

Validation of normalised velocity deficit predictions obtained with the proposed analytical wake models are presented in Fig. \ref{fig:cases1profiles} with horizontal ($y$) and vertical ($z$) profiles across the turbine wake centre at downstream distances of $x/D$ = 3, 6, 9 and 12. 
For completeness, we include the LES-ALM results from \citet{Abkar2017}.
Our Gaussian model provides a similar accuracy to that of \citet{Abkar2019} at most profiles, with ours with an slight underprediction of $\Delta U/U_0$ for $x/D=3$ observed in both vertical and horizontal profiles. 
Note that the model from \citet{Abkar2019} slightly overpredicts the LES-ALM velocity deficit values at the wake centre at distances of $x/D=9$ and 12, whereas ours attains a closer match to the LES data.  

\begin{figure}
\centerline{\includegraphics[width=0.8\linewidth]{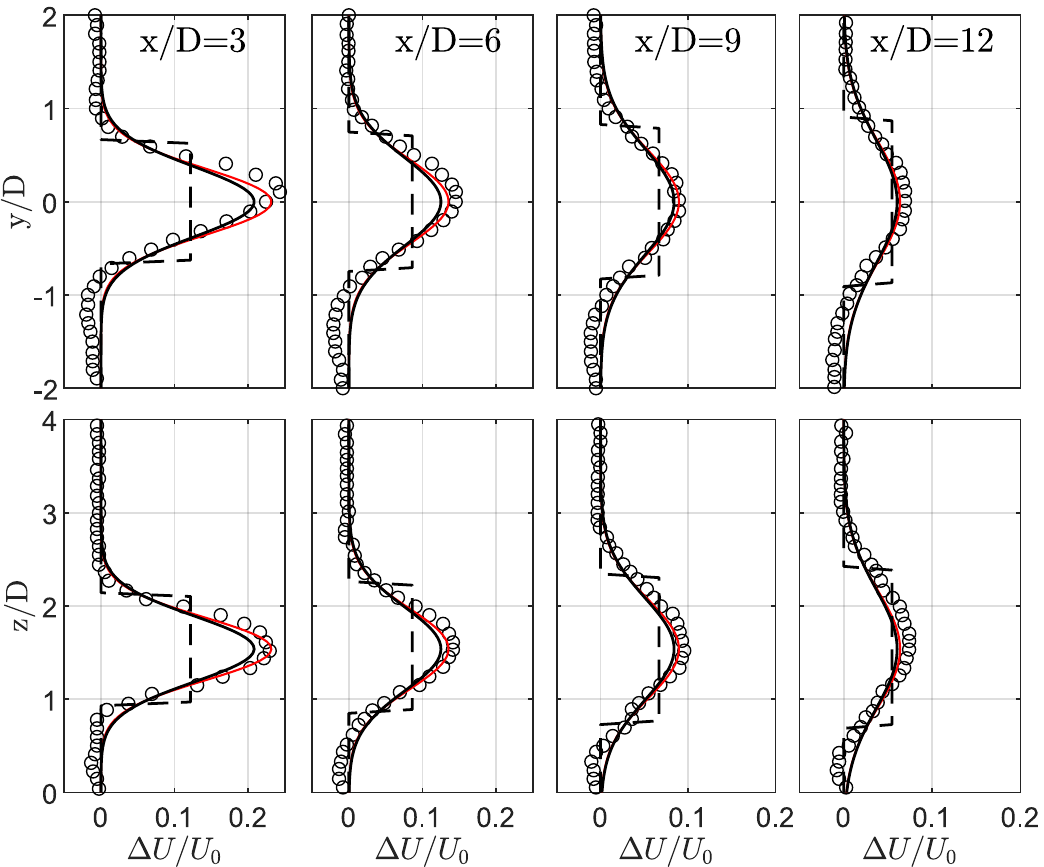}}
\caption{Normalised velocity deficit profiles for case 1b with $C_T$ = 0.34, TSR = 2.5, $D$ = 26 m and $H$ = 24 m. Same legend as Fig. \ref{fig:cases1profiles}.}  \label{fig:cases2profiles}
\end{figure}

Figure \ref{fig:cases2profiles} presents the results obtained for case 1b in which the VAWT operates at a lower TSR which decreases its $C_T$ in comparison to case 1a (and case 1c). 
We observe our Gaussian wake model provides again a good estimate of the maximum velocity deficit at the wake centre over the vertical and horizontal directions, and the wake width is also in agreement with the LES and wake model from \citet{Abkar2019}.
As VAWT wakes asymmetry depends on its aspect ratio \citep{Shamsoddin2020}, in cases 1a and 1b this is approx. unity leading the wake to feature a Gaussian distribution over the horizontal and vertical directions at $x/D \geq 3$, which supports the self-similarity assumption.
Some degree of asymmetry in $\Delta U/U_0$ is appreciated at $x/D=3$ with its maximum value over the horizontal plane being slightly larger than in the vertical one.

\begin{figure}
\centerline{\includegraphics[width=0.8\linewidth]{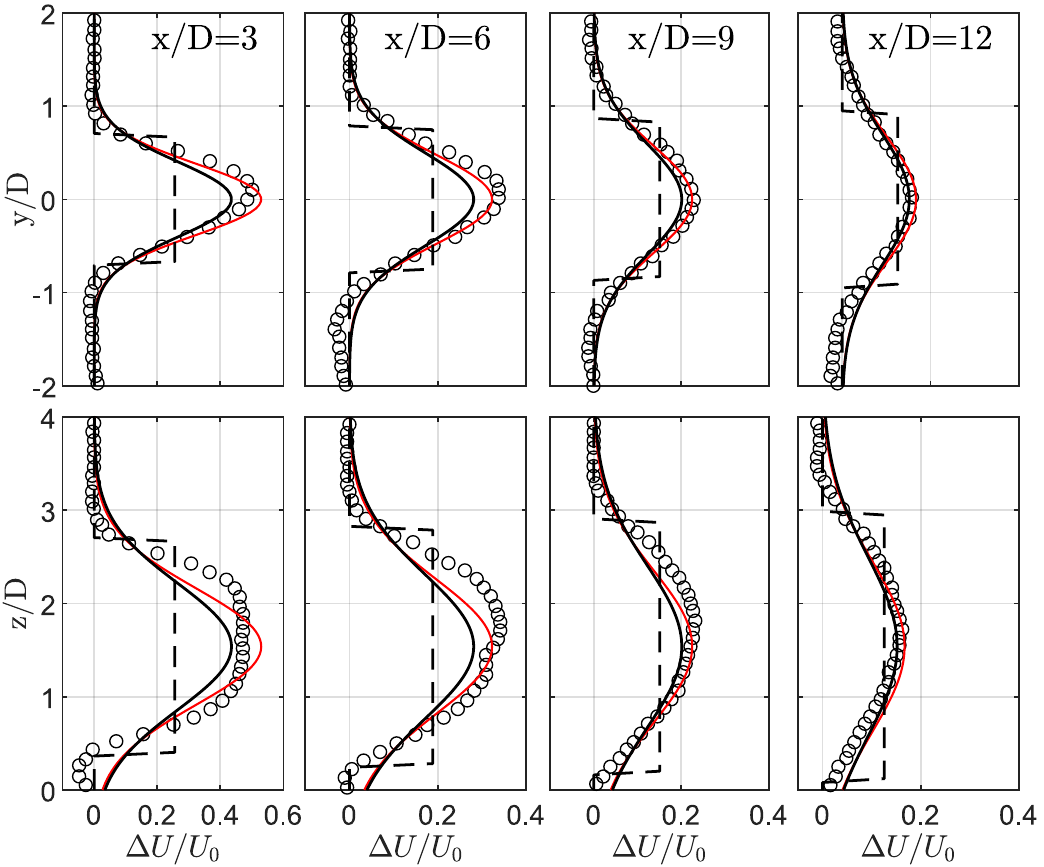}}
\caption{Normalised velocity deficit profiles for case 1c with $C_T$ = 0.63, TSR = 3.8, $D$ = 26 m and $H$ = 48 m. Same legend as in Fig. \ref{fig:cases1profiles}.}  \label{fig:cases3profiles}
\end{figure}

The turbine in case 1c has an aspect ratio close to two that promotes an uneven recovery over the vertical and horizontal directions which is well observed in Fig. \ref{fig:cases3profiles} with the distribution of $\Delta U/U_0$ over the $y$-direction profiles attaining a Gaussian profile whilst, over the vertical direction, the profiles nearer to the VAWT exhibit an almost top-hat distribution.
After $x/D=9$ which corresponds to $x/H=4.5$, the LES data indicate the velocity deficit in the VAWT wake over the vertical direction recovers the Gaussian distribution.
Our Gaussian wake model provides a satisfactory representation of $\Delta U/U_0$ as at $x/D=3$ its maximum value is closer to the LES data than the one predicted by the model from \citet{Abkar2019} which overpredicts this value.
Far downstream of the turbine, at $x/D$ = 9 and 12, our wake model achieves a good agreement with the LES results for the maximum velocity deficit.

\begin{figure}
\centerline{\includegraphics[width=0.9\linewidth]{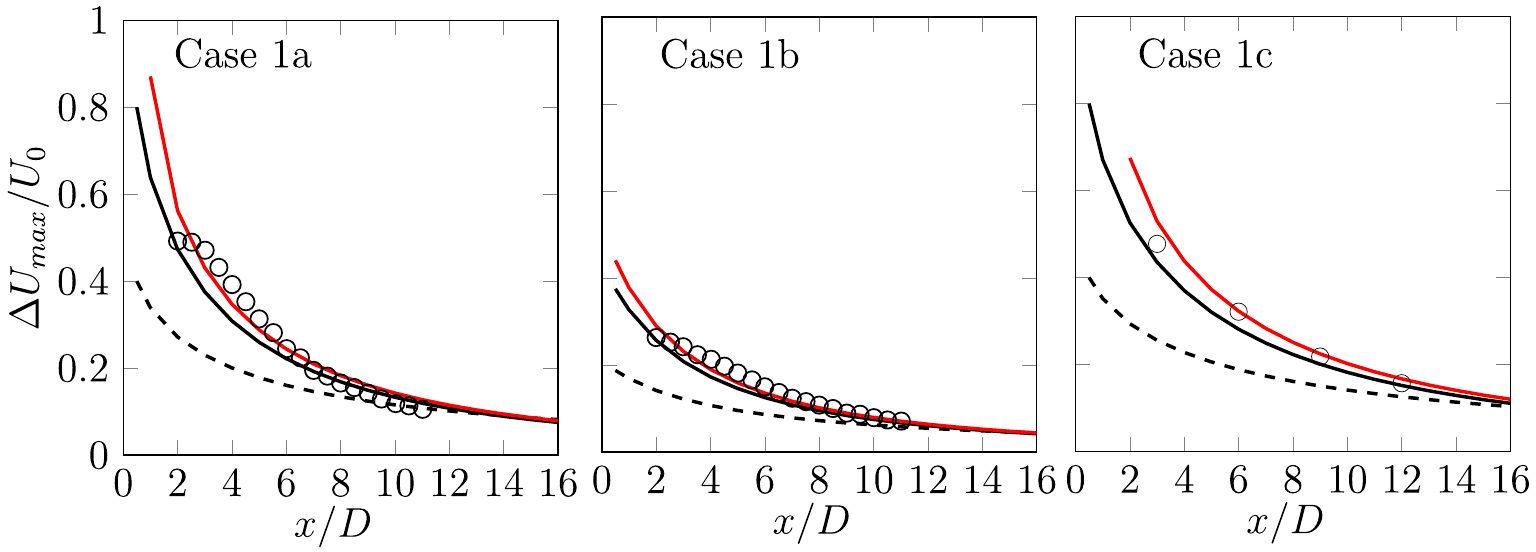}}
\caption{Normalised maximum velocity deficit for cases 1a to 1c. Same legend as in Fig. \ref{fig:cases1profiles}.}  \label{fig:cases1to3_DUmax}
\end{figure}

Further analysis of the wake predictions is presented in Fig. \ref{fig:cases1to3_DUmax} with the evolution of maximum velocity deficit ($\Delta U_{max}/U_0$) over the streamwise direction for cases 1a to 1c, which represents the velocity scale adopted in the wake model adopted in the Gaussian and top-hat velocity distributions.
In comparison to the LES results, our Gaussian model estimates a faster velocity recovery, i.e. lower maximum wake velocity deficit, than that from \citet{Abkar2019}, which is a result of the considered rectangular wake onset area at $x_a$ represented by $\varepsilon_y$ and $\varepsilon_z$.
For cases 1a and 1b, all analytical models predict similar values of $\Delta U_{max}/U_0$ for distances $x/D_0 \geq 8$.
Slightly larger differences are observed for case 1c in which the wake is more asymmetric, with our Gaussian analytical wake model providing a closer prediction to those from the LES.
We also note that the latter is able to provide wake velocity values at any streamwise distance even immediately downstream of the VAWT, whilst \citet{Abkar2019} model fails at providing physical values for $x/D \leq 2.0$ in case 1c due to determining the value of $\beta$ partly based on HAT wake characteristics.

\begin{figure}
\centerline{\includegraphics[width=0.9\linewidth]{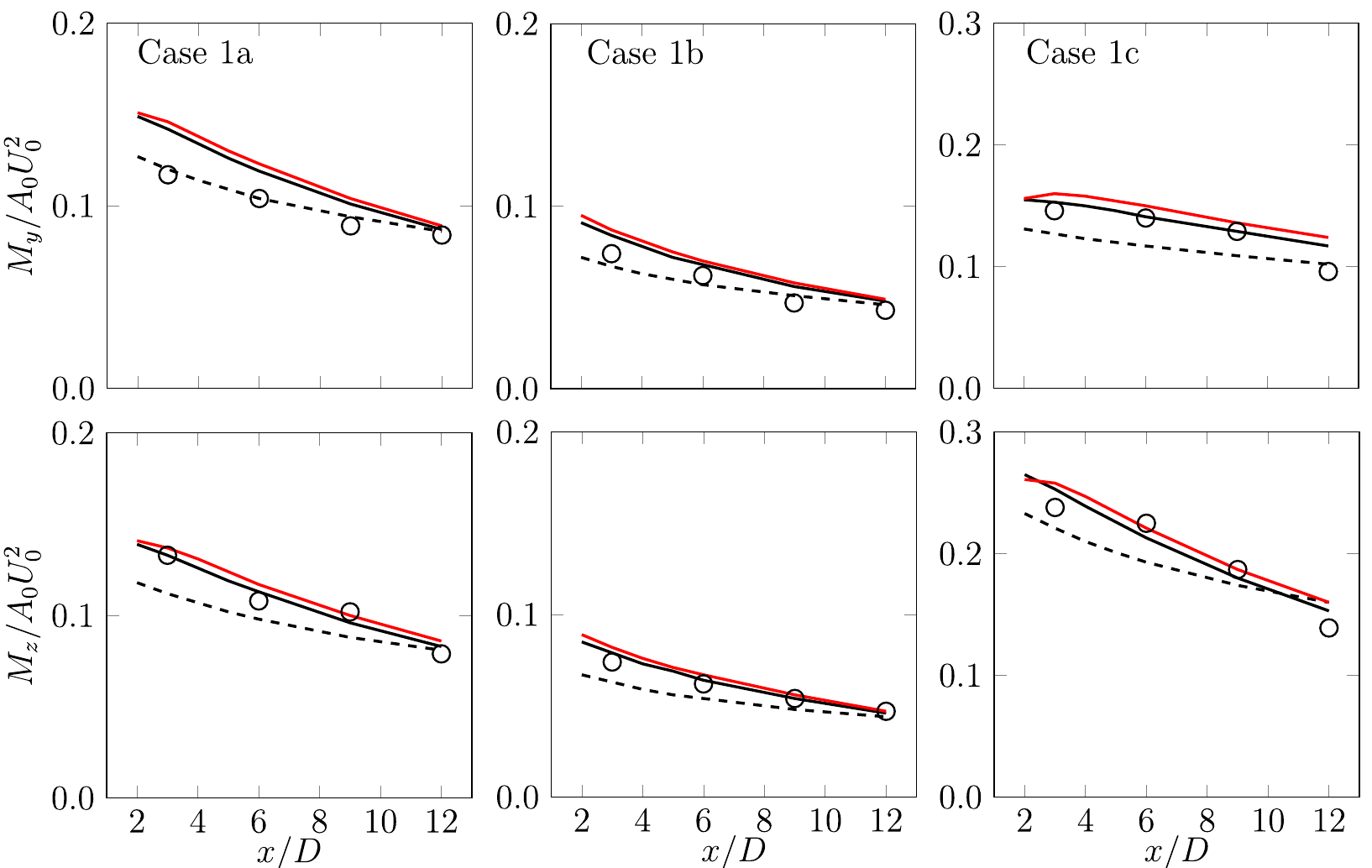}}
\caption{Predictions of momentum thickness for cases 1a, 1b and 1c. Same legend as in Fig. \ref{fig:cases1profiles}.}  \label{fig:case1MomThick}
\end{figure}

In turbine array modelling, any analytical wake model needs to provide an accurate downstream evolution of the wake deficit so wake-to-wake interactions are reliably accounted for, which is key to estimate the performance of the individual devices as power scales according to $P \propto U_{0_i}^3$.
Fig. \ref{fig:case1MomThick} presents the streamwise evolution of the normalised total momentum deficit at the centre planes of the wake, i.e. computed over the horizontal ($M_y$) and vertical ($M_z$) directions at $z=0$ and $y=0$ respectively, which result from combining \ref{eq:Thrust1} and \ref{Thrust2}, as

\begin{equation}
M_{i} = 
\frac{2 \int U_{w}\left(U_{0} - U_{w}\right)\mathrm{d} x_i}{A_0 U_{0}^2} \label{eq:momdef}
\end{equation}

An overall good agreement of the total momentum deficit is observed with the Gaussian models providing a closer match with the LES data, except for $M_y$ in case 1a with our top-hat model attains a better match.
In line with the maximum velocity deficit results, our Gaussian model predicts lower total momentum deficit which appear closer to the LES data compared to the estimates from the other theoretical models.
LES results from case 1c depict the asymmetric wake distribution with larger values for $M_z$ than $M_y$, with the latter decaying at a slower rate than the vertical momentum deficit, which is well predicted by all wake models.
The top-hat model appears to consistently underestimate the value of momentum deficits in most cases. 

\subsection{Array of four aligned VAWTs}

The largest experimental facility with full-scale VAWTs, FLOWE (US), has 18 devices rated at 1.2MW and research undertaken at this site enabled the world-first quantification of a VAWT array performance and flow mechanisms driving the wake recovery \citep{Kinzel2012}.
To quantify the accuracy of our analytical models, we consider a configuration with four VAWTs fully aligned with the onset flow, i.e. the second to fourth turbines operate in complete waked conditions, with a separation of $11D$ between them.
The turbines have a diameter and height of 1.2 m and 6.1 m respectively, with a separation of 3 m from the bottom tip to the ground level, i.e. a total height of 9.1 m, and operate at TSR = 2.3. 
The free-stream velocity is 8.45 ms$^{-1}$ with a turbulence intensity of 11\% considered for all turbines, and flow statistics in the wakes are obtained at seven equally-spaced bins that span over the whole turbine height, whose measures are then integrated into a single vertically averaged value.
We estimate a thrust coefficient of 0.652 based on the field power coefficient of 0.134 calculated in \citet{Kinzel2012} based on the free-stream velocity, using the induction factor of the actuator disk theory.

\begin{figure}
\centerline{\includegraphics[width=0.9\linewidth]{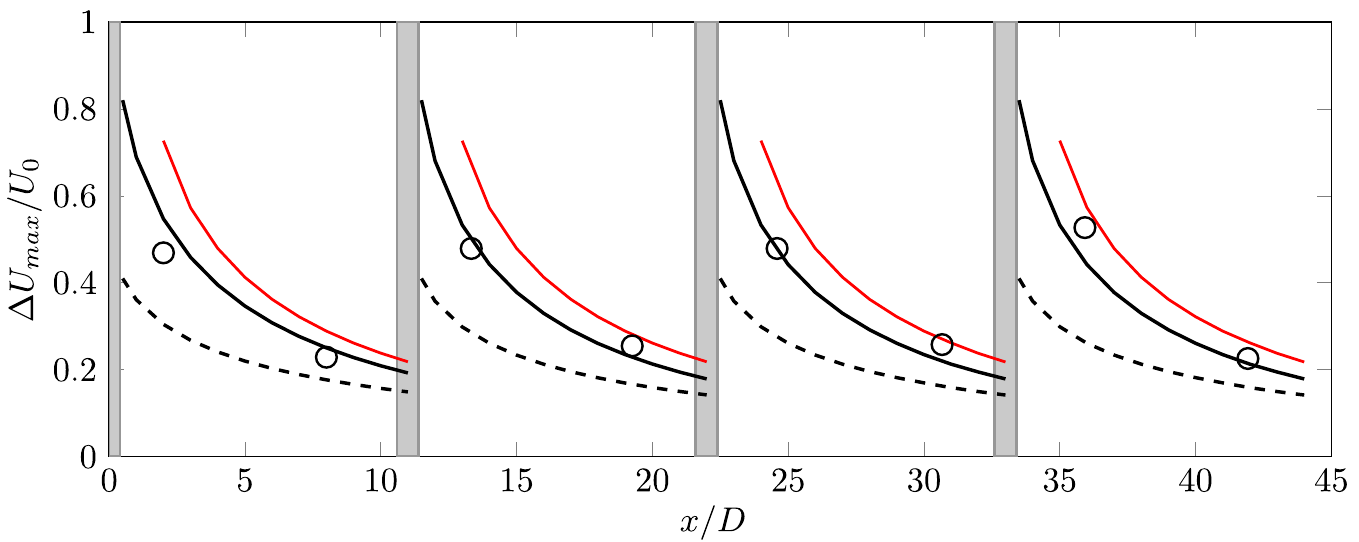}}
\caption{Normalised maximum velocity deficit profiles for the case with four full-scale VAWT. Comparison of our proposed top-hat (dashed line) and Gaussian (solid black line) analytical wake models, with the Gaussian model proposed by \citet{Abkar2019} (solid red line) and field data from \citet{Kinzel2012} (circles).}  \label{fig:case5}
\end{figure}

Comparison of the maximum velocity deficit over the centre-line of the turbines is presented in Fig. \ref{fig:case5}, including the field data which measured velocities at 2$D$ and 8$D$ downstream of each device, and our two proposed wake models with that from \citet{Abkar2019}.
Our Gaussian wake model provides a good accuracy when compared to the field data for those points further downstream of each turbine whilst slightly over predicts the maximum velocity deficit at a distance of $2D$ downstream of the first and fourth turbine. 
The latter could be attributed to the wake velocity distribution not featuring full self-similarity in the wake profiles at such short distance given the aspect ratio of these VAWTs is almost of five. 
These results evidence the relevance of the initial wake expansion correctly accounted for in our Gaussian model compared to the one proposed in \citet{Abkar2019} which overestimates the wake velocity deficit throughout the array.
Our top-hat model notably underestimates the maximum velocity deficit, in line with the results presented in \S\ref{sec:cases1}, but these proved useful in the computation of the momentum thickness in the far-wake, thus somewhat valuable to predict the performance of the turbines in arrays.

\begin{figure}
\centerline{\includegraphics[width=0.9\linewidth]{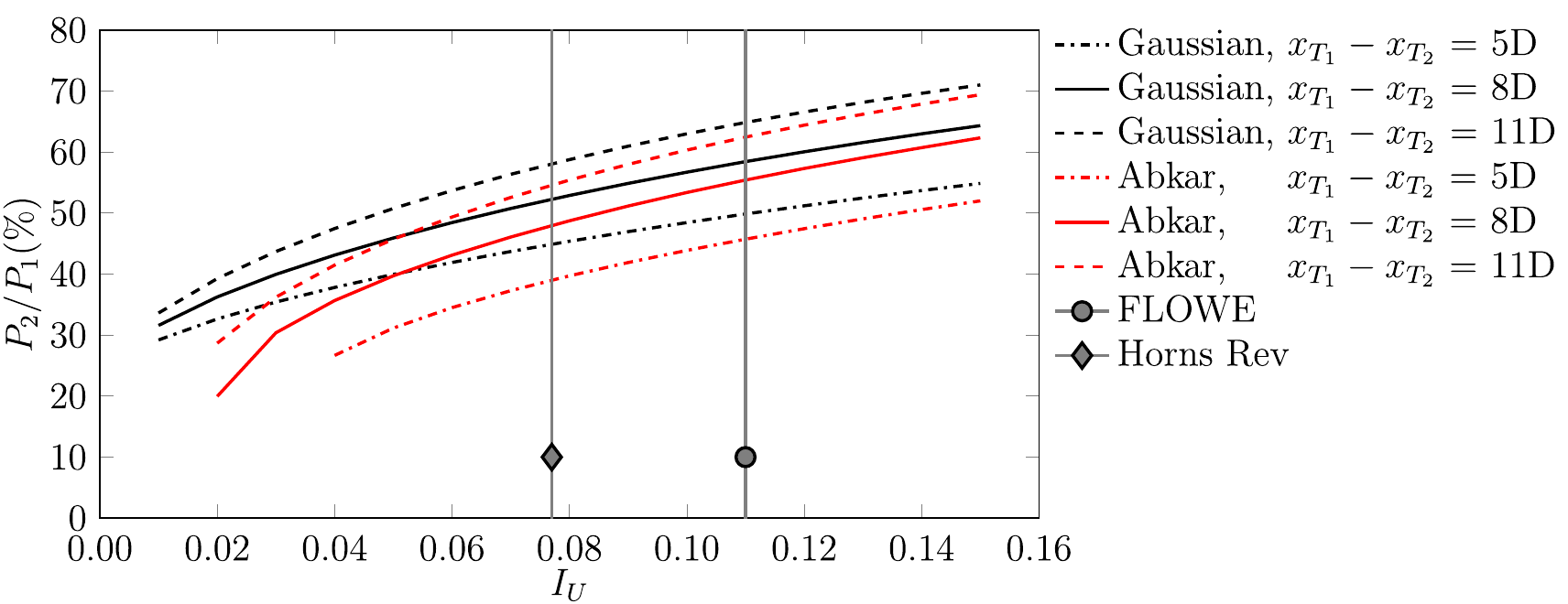}}
\caption{Evolution of the power ratio between the first two turbines at FLOWE for different levels of turbulence intensity. Comparison between our proposed Gaussian analytical wake model (black line) and that of \citet{Abkar2019} (red line).}  \label{fig:TI}
\end{figure}

A key feature of analytical wake models is that they need to provide reliable power estimates of turbines in arrays, being this dependent on the velocity field such as $P \propto U^3$.
In Fig. \ref{fig:TI} we provide a quantitative comparison of the ratio of power generated by the first turbine $T_1$, assuming $U_{T_1}=U_0$, and the turbine $T_2$ immediately behind, with $U_{T_2}$ = $U_w(x)$.
The power ratio $P_1/P_2$ is estimated for various distances between these turbines, $x_{T_1}-x_{T_2}$, considering $11D$ as in FLOWE, $8D$ and $5D$, and over a range of turbulence intensities $0.01 < I_u < 0.15$ to consider scenarios representative of offshore ($I_u$ = 6--8\%) or onshore ($I_u$ = 10--12\%) locations \citep{Barthelmie2007}.
Results show power production from the downstream turbine obtained with our Gaussian model is consistently larger than with Abkar's model, in line with the evolution of velocity deficit seen in Fig. \ref{fig:case5}.
Considering the well-known HAWT array of Horns Rev which can feature turbulence intensities of approx. 7--8\%, with turbine spacing of $11D$, $8D$ and $5D$ the differences in predictions of $P_1/P_2$ are 6\%, 9\% and 14\% respectively, with Abkar's model underestimating the values as it accounts for larger velocity deficit as shown in Fig. \ref{fig:case5}.  
At high turbulence intensity values, variation in power estimates between models narrows down as the wake expansion increases with $I_u$ leading to faster momentum recovery, in turn diminishing the relevance of the onset wake width prediction (Eq. \ref{eq:varepsVAT}).
Overall, this power-prediction sensitivity shows that the presented improvements in our Gaussian model can avoid an underestimation of the array efficiency, especially for environments with low turbulence levels.

\section{Conclusions} \label{sec:conclusions}

This paper presents a new set of top-hat and Gaussian wake models for Vertical Axis Turbines (VATs) built upon considering a rectangular wake cross-section that evolves according to two length-scales and including an asymmetric expansion over the horizontal and vertical planes.
In comparison to previous wake models, we included three main features: 
the location at which the wake pressure is balanced is attained at $x_a=D/2$, with $D$ being the turbine diameter; 
the wake at this location has a rectangular cross-section as the VAT rotor projected area that scales with a factor $\varepsilon = (\beta/4 \pi)^{1/2}$; and,
the asymmetric wake recovery dynamics is represented by uneven wake expansion rates.

We validated the accuracy of our proposed models in comparison to large-eddy simulation results and field data in four cases involving a standalone and an array of vertical axis wind turbines.
We proved the suitability of our models to estimate the wake velocities for all cases at any downstream location, providing improvements compared to other existing VAT wake models.
Our analytical models provide a good estimation of the maximum velocity deficit and momentum thickness for the three cases with a single turbine that operated at two tip-speed ratios and had two diameter-to-height aspect ratios. 
In application to an array of four full-scale turbines, the models compared well with field data with results showing that our Gaussian wake model attains a good estimation of the maximum velocity deficit, which verifies its reliability to predict the interaction between turbines in arrays.
In relation to power generation, we performed a sensitivity analysis over a range of turbulence intensity values and turbine spacing showing there is a larger deviation between models for low turbulence values, with our Gaussian model providing larger power generation values due to its better prediction of the velocity deficit field.

Overall, the results confirmed an enhanced accuracy of our wake models compared to other previously proposed models, which will lead to improved designs of VAT arrays increasing the development pace of this promising technology.


\section*{Acknowledgements}
This research was partially supported by the UK's Engineering and Physical Sciences Research Council (EPSRC) through the project EP/R51150X/1.

\bibliographystyle{model1-num-names}
\bibliography{VATbib.bib}

\begin{thebibliography}{33}
\expandafter\ifx\csname natexlab\endcsname\relax\def\natexlab#1{#1}\fi
\providecommand{\bibinfo}[2]{#2}
\ifx\xfnm\relax \def\xfnm[#1]{\unskip,\space#1}\fi
\bibitem[{Veers et~al.(2019)Veers, Dykes, Lantz, Barth, Bottasso, Carlson,
  Clifton, Green, Green, Holttinen, Laird, Lehtomäki, Lundquist, Manwell,
  Marquis, Meneveau, Moriarty, Munduate, Muskulus, Naughton, Pao, Paquette,
  Peinke, Robertson, Rodrigo, Sempreviva, Smith, Tuohy, and Wiser}]{Veers2019}
\bibinfo{author}{P.~Veers}, \bibinfo{author}{K.~Dykes},
  \bibinfo{author}{E.~Lantz}, \bibinfo{author}{S.~Barth},
  \bibinfo{author}{C.~L. Bottasso}, \bibinfo{author}{O.~Carlson},
  \bibinfo{author}{A.~Clifton}, \bibinfo{author}{J.~Green},
  \bibinfo{author}{P.~Green}, \bibinfo{author}{H.~Holttinen},
  \bibinfo{author}{D.~Laird}, \bibinfo{author}{V.~Lehtomäki},
  \bibinfo{author}{J.~K. Lundquist}, \bibinfo{author}{J.~Manwell},
  \bibinfo{author}{M.~Marquis}, \bibinfo{author}{C.~Meneveau},
  \bibinfo{author}{P.~Moriarty}, \bibinfo{author}{X.~Munduate},
  \bibinfo{author}{M.~Muskulus}, \bibinfo{author}{J.~Naughton},
  \bibinfo{author}{L.~Pao}, \bibinfo{author}{J.~Paquette},
  \bibinfo{author}{J.~Peinke}, \bibinfo{author}{A.~Robertson},
  \bibinfo{author}{J.~S. Rodrigo}, \bibinfo{author}{A.~M. Sempreviva},
  \bibinfo{author}{J.~C. Smith}, \bibinfo{author}{A.~Tuohy},
  \bibinfo{author}{R.~Wiser},
\newblock \bibinfo{title}{{Grand challenges in the science of wind energy}},
\newblock \bibinfo{journal}{Science} \bibinfo{volume}{366}
  (\bibinfo{year}{2019}) \bibinfo{pages}{4443}.
\bibitem[{Liu et~al.(2019)Liu, Lin, and Zhang}]{Liu2019}
\bibinfo{author}{J.~Liu}, \bibinfo{author}{H.~Lin}, \bibinfo{author}{J.~Zhang},
\newblock \bibinfo{title}{{Review on the technical perspectives and commercial
  viability of vertical axis wind turbines}},
\newblock \bibinfo{journal}{Ocean Engineering} \bibinfo{volume}{182}
  (\bibinfo{year}{2019}) \bibinfo{pages}{608--626}.
\bibitem[{Rolin and Port{\'{e}}-Agel(2018)}]{Rolin2018}
\bibinfo{author}{V.~Rolin}, \bibinfo{author}{F.~Port{\'{e}}-Agel},
\newblock \bibinfo{title}{{Experimental investigation of vertical-axis
  wind-turbine wakes in boundary layer flow}},
\newblock \bibinfo{journal}{Renewable Energy} \bibinfo{volume}{118}
  (\bibinfo{year}{2018}) \bibinfo{pages}{1--13}.
\bibitem[{Araya et~al.(2017)Araya, Colonius, and Dabiri}]{Araya2017}
\bibinfo{author}{D.~Araya}, \bibinfo{author}{T.~Colonius},
  \bibinfo{author}{J.~O. Dabiri},
\newblock \bibinfo{title}{{Transition to bluff-body dynamics in the wake of
  vertical-axis wind turbines}},
\newblock \bibinfo{journal}{Journal of Fluid Mechanics} \bibinfo{volume}{813}
  (\bibinfo{year}{2017}) \bibinfo{pages}{346--381}.
\bibitem[{Ouro et~al.(2019)Ouro, Runge, Luo, and Stoesser}]{Ouro2019RE}
\bibinfo{author}{P.~Ouro}, \bibinfo{author}{S.~Runge},
  \bibinfo{author}{Q.~Luo}, \bibinfo{author}{T.~Stoesser},
\newblock \bibinfo{title}{{Three-dimensionality of the wake recovery behind a
  vertical axis turbine}},
\newblock \bibinfo{journal}{Renewable Energy} \bibinfo{volume}{133}
  (\bibinfo{year}{2019}) \bibinfo{pages}{1066--1077}.
\bibitem[{Parker and Leftwich(2016)}]{Parker2016}
\bibinfo{author}{C.~M. Parker}, \bibinfo{author}{M.~C. Leftwich},
\newblock \bibinfo{title}{{The effect of tip speed ratio on a vertical axis
  wind turbine at high Reynolds numbers}},
\newblock \bibinfo{journal}{Experiments in Fluids} \bibinfo{volume}{57}
  (\bibinfo{year}{2016}) \bibinfo{pages}{74}.
\bibitem[{Posa(2020)}]{Posa2020a}
\bibinfo{author}{A.~Posa},
\newblock \bibinfo{title}{{Influence of Tip Speed Ratio on wake features of a
  Vertical Axis Wind Turbine}},
\newblock \bibinfo{journal}{Journal of Wind Engineering and Industrial
  Aerodynamics} \bibinfo{volume}{197} (\bibinfo{year}{2020})
  \bibinfo{pages}{104076}.
\bibitem[{Ouro and Stoesser(2017)}]{Ouro2017CAF}
\bibinfo{author}{P.~Ouro}, \bibinfo{author}{T.~Stoesser},
\newblock \bibinfo{title}{{An immersed boundary-based large-eddy simulation
  approach to predict the performance of vertical axis tidal turbines}},
\newblock \bibinfo{journal}{Computers {\&} Fluids} \bibinfo{volume}{152}
  (\bibinfo{year}{2017}) \bibinfo{pages}{74--87}.
\bibitem[{Tescione et~al.(2014)Tescione, Ragni, He, {Sim{\~{a}}o Ferreira}, and
  van Bussel}]{Tescione2014}
\bibinfo{author}{G.~Tescione}, \bibinfo{author}{D.~Ragni},
  \bibinfo{author}{C.~He}, \bibinfo{author}{C.~J. {Sim{\~{a}}o Ferreira}},
  \bibinfo{author}{G.~van Bussel},
\newblock \bibinfo{title}{{Near wake flow analysis of a vertical axis wind
  turbine by stereoscopic particle image velocimetry}},
\newblock \bibinfo{journal}{Renewable Energy} \bibinfo{volume}{70}
  (\bibinfo{year}{2014}) \bibinfo{pages}{47--61}.
\bibitem[{Brochier et~al.(1986)Brochier, Fraunie, Beguier, and
  Paraschivoiu}]{Brochier1986}
\bibinfo{author}{G.~Brochier}, \bibinfo{author}{P.~Fraunie},
  \bibinfo{author}{C.~Beguier}, \bibinfo{author}{I.~Paraschivoiu},
\newblock \bibinfo{title}{{Water channel experiments of dynamic stall on
  Darrieus wind turbine blades}},
\newblock \bibinfo{journal}{Journal of Propulsion} \bibinfo{volume}{2}
  (\bibinfo{year}{1986}) \bibinfo{pages}{445--449}.
\bibitem[{Bachant and Wosnik(2016)}]{Bachant2016}
\bibinfo{author}{P.~Bachant}, \bibinfo{author}{M.~Wosnik},
\newblock \bibinfo{title}{{Effects of Reynolds Number on the Energy Conversion
  and Near-Wake Dynamics of a High Solidity Vertical-Axis Cross-Flow Turbine}},
\newblock \bibinfo{journal}{Energies} \bibinfo{volume}{9}
  (\bibinfo{year}{2016}) \bibinfo{pages}{73}.
\bibitem[{Bastankhah and Port{\'{e}}-Agel(2014)}]{Bastankhah2014}
\bibinfo{author}{M.~Bastankhah}, \bibinfo{author}{F.~Port{\'{e}}-Agel},
\newblock \bibinfo{title}{{A new analytical model for wind-turbine wakes}},
\newblock \bibinfo{journal}{Renewable Energy} \bibinfo{volume}{70}
  (\bibinfo{year}{2014}) \bibinfo{pages}{116--123}.
\bibitem[{Kadum et~al.(2020)Kadum, Bayo\'an~Cal, Quigley, Cortina, and
  Calaf}]{Kadum2020}
\bibinfo{author}{H.~Kadum}, \bibinfo{author}{R.~Bayo\'an~Cal},
  \bibinfo{author}{M.~Quigley}, \bibinfo{author}{G.~Cortina},
  \bibinfo{author}{M.~Calaf},
\newblock \bibinfo{title}{{Compounded energy gains in collocated wind plants:
  Energy balance quantification and wake morphology description}},
\newblock \bibinfo{journal}{Renewable Energy} \bibinfo{volume}{150}
  (\bibinfo{year}{2020}) \bibinfo{pages}{868--877}.
\bibitem[{Shamsoddin and Port{\'{e}}-Agel(2020)}]{Shamsoddin2020}
\bibinfo{author}{S.~Shamsoddin}, \bibinfo{author}{F.~Port{\'{e}}-Agel},
\newblock \bibinfo{title}{{Effect of aspect ratio on vertical-axis wind turbine
  wakes}},
\newblock \bibinfo{journal}{Journal of Fluid Mechanics} \bibinfo{volume}{889}
  (\bibinfo{year}{2020}) \bibinfo{pages}{R1}.
\bibitem[{Ouro et~al.(2018)Ouro, Stoesser, and Ram\'irez}]{Ouro2018JFE}
\bibinfo{author}{P.~Ouro}, \bibinfo{author}{T.~Stoesser},
  \bibinfo{author}{L.~Ram\'irez},
\newblock \bibinfo{title}{{Effect of Blade Cambering on Dynamic Stall in View
  of Designing Vertical Axis Turbines}},
\newblock \bibinfo{journal}{ASME Journal of Fluids Engineering}
  \bibinfo{volume}{140} (\bibinfo{year}{2018}) \bibinfo{pages}{061104}.
\bibitem[{Posa(2020)}]{Posa2020b}
\bibinfo{author}{A.~Posa},
\newblock \bibinfo{title}{{Dependence of the wake recovery downstream of a
  Vertical Axis Wind Turbine on its dynamic solidity}},
\newblock \bibinfo{journal}{Journal of Wind Engineering and Industrial
  Aerodynamics} \bibinfo{volume}{202} (\bibinfo{year}{2020})
  \bibinfo{pages}{104212}.
\bibitem[{Posa and Balaras(2018)}]{Posa2018}
\bibinfo{author}{A.~Posa}, \bibinfo{author}{E.~Balaras},
\newblock \bibinfo{title}{{Large Eddy Simulation of an isolated vertical axis
  wind turbine}},
\newblock \bibinfo{journal}{Journal of Wind Engineering and Industrial
  Aerodynamics} \bibinfo{volume}{172} (\bibinfo{year}{2018})
  \bibinfo{pages}{139--151}.
\bibitem[{Port{\'{e}}-Agel et~al.(2020)Port{\'{e}}-Agel, Bastankhah, and
  Shamsoddin}]{Porte-Agel2020}
\bibinfo{author}{F.~Port{\'{e}}-Agel}, \bibinfo{author}{M.~Bastankhah},
  \bibinfo{author}{S.~Shamsoddin},
\newblock \bibinfo{title}{{Wind-Turbine and Wind-Farm Flows: A Review}},
\newblock \bibinfo{journal}{Boundary-Layer Meteorology} \bibinfo{volume}{174}
  (\bibinfo{year}{2020}) \bibinfo{pages}{1--59}.
\bibitem[{Abkar and Dabiri(2017)}]{Abkar2017}
\bibinfo{author}{M.~Abkar}, \bibinfo{author}{J.~O. Dabiri},
\newblock \bibinfo{title}{{Self-similarity and flow characteristics of
  vertical-axis wind turbine wakes: an LES study}},
\newblock \bibinfo{journal}{Journal of Turbulence} \bibinfo{volume}{18}
  (\bibinfo{year}{2017}) \bibinfo{pages}{373--389}.
\bibitem[{Shamsoddin and Port{\'{e}}-Agel(2014)}]{Shamsoddin2014}
\bibinfo{author}{S.~Shamsoddin}, \bibinfo{author}{F.~Port{\'{e}}-Agel},
\newblock \bibinfo{title}{{Large eddy simulation of vertical axis wind turbine
  wakes}},
\newblock \bibinfo{journal}{Energies} \bibinfo{volume}{7}
  (\bibinfo{year}{2014}) \bibinfo{pages}{890--912}.
\bibitem[{Shamsoddin and Port{\'{e}}-Agel(2016)}]{Shamsoddin2016}
\bibinfo{author}{S.~Shamsoddin}, \bibinfo{author}{F.~Port{\'{e}}-Agel},
\newblock \bibinfo{title}{{A Large-Eddy Simulation Study of Vertical Axis Wind
  Turbine Wakes in the Atmospheric Boundary Layer}},
\newblock \bibinfo{journal}{Energies} \bibinfo{volume}{9}
  (\bibinfo{year}{2016}) \bibinfo{pages}{366}.
\bibitem[{Massie et~al.(2019)Massie, Ouro, Stoesser, and Luo}]{Massie2019}
\bibinfo{author}{L.~Massie}, \bibinfo{author}{P.~Ouro},
  \bibinfo{author}{T.~Stoesser}, \bibinfo{author}{Q.~Luo},
\newblock \bibinfo{title}{{An Actuator Surface Model to Simulate Vertical Axis
  Turbines}},
\newblock \bibinfo{journal}{Energies} \bibinfo{volume}{12}
  (\bibinfo{year}{2019}) \bibinfo{pages}{4741}.
\bibitem[{Meneveau(2019)}]{Meneveau2019}
\bibinfo{author}{C.~Meneveau},
\newblock \bibinfo{title}{{Big wind power: seven questions for turbulence
  research}},
\newblock \bibinfo{journal}{Journal of Turbulence} \bibinfo{volume}{20}
  (\bibinfo{year}{2019}) \bibinfo{pages}{2--20}.
\bibitem[{Stansby and Stallard(2016)}]{Stansby2016}
\bibinfo{author}{P.~K. Stansby}, \bibinfo{author}{T.~Stallard},
\newblock \bibinfo{title}{{Fast optimisation of tidal stream turbine positions
  for power generation in small arrays with low blockage based on superposition
  of self-similar far-wake velocity deficit profiles}},
\newblock \bibinfo{journal}{Renewable Energy} \bibinfo{volume}{92}
  (\bibinfo{year}{2016}) \bibinfo{pages}{366--375}.
\bibitem[{Stevens and Meneveau(2017)}]{Stevens2017}
\bibinfo{author}{R.~J. Stevens}, \bibinfo{author}{C.~Meneveau},
\newblock \bibinfo{title}{{Flow Structure and Turbulence in Wind Farms}},
\newblock \bibinfo{journal}{Annual Review of Fluid Mechanics}
  \bibinfo{volume}{49} (\bibinfo{year}{2017}) \bibinfo{pages}{311--339}.
\bibitem[{Frandsen et~al.(2006)Frandsen, Barthelmie, Pryor, Rathmann, Larsen,
  H{\o}jstrup, and Th{\o}gersen}]{Frandsen2006}
\bibinfo{author}{S.~Frandsen}, \bibinfo{author}{R.~Barthelmie},
  \bibinfo{author}{S.~Pryor}, \bibinfo{author}{O.~Rathmann},
  \bibinfo{author}{S.~Larsen}, \bibinfo{author}{J.~H{\o}jstrup},
  \bibinfo{author}{M.~Th{\o}gersen},
\newblock \bibinfo{title}{{Analytical modelling of wind speed deficit in large
  offshore wind farms}},
\newblock \bibinfo{journal}{Wind Energy} \bibinfo{volume}{9}
  (\bibinfo{year}{2006}) \bibinfo{pages}{39--53}.
\bibitem[{Tennekes and Lumley(1972)}]{Tennekes}
\bibinfo{author}{H.~Tennekes}, \bibinfo{author}{J.~Lumley}, \bibinfo{title}{{An
  First Course in Turbulence}}, \bibinfo{publisher}{The MIT press},
  \bibinfo{year}{1972}.
\bibitem[{Abkar(2019)}]{Abkar2019}
\bibinfo{author}{M.~Abkar},
\newblock \bibinfo{title}{{Theoretical modeling of vertical-axis wind turbine
  wakes}},
\newblock \bibinfo{journal}{Energies} \bibinfo{volume}{12}
  (\bibinfo{year}{2019}).
\bibitem[{Stallard et~al.(2015)Stallard, Feng, and Stansby}]{Stallard2015}
\bibinfo{author}{T.~Stallard}, \bibinfo{author}{T.~Feng},
  \bibinfo{author}{P.~K. Stansby},
\newblock \bibinfo{title}{{Experimental study of the mean wake of a tidal
  stream rotor in a shallow turbulent flow}},
\newblock \bibinfo{journal}{Journal of Fluids and Structures}
  \bibinfo{volume}{54} (\bibinfo{year}{2015}) \bibinfo{pages}{235--246}.
\bibitem[{Bastankhah and Port{\'{e}}-Agel(2016)}]{Bastankhah2016}
\bibinfo{author}{M.~Bastankhah}, \bibinfo{author}{F.~Port{\'{e}}-Agel},
\newblock \bibinfo{title}{{Experimental and theoretical study of wind turbine
  wakes in yawed conditions}},
\newblock \bibinfo{journal}{Journal of Fluid Mechanics} \bibinfo{volume}{806}
  (\bibinfo{year}{2016}) \bibinfo{pages}{506--541}.
\bibitem[{Shapiro et~al.(2018)Shapiro, Gayme, and Meneveau}]{Shapiro2018}
\bibinfo{author}{C.~R. Shapiro}, \bibinfo{author}{D.~F. Gayme},
  \bibinfo{author}{C.~Meneveau},
\newblock \bibinfo{title}{{Modelling yawed wind turbine wakes: a lifting line
  approach}},
\newblock \bibinfo{journal}{Journal of Fluid Mechanics} \bibinfo{volume}{841}
  (\bibinfo{year}{2018}) \bibinfo{pages}{R1}.
\bibitem[{Kinzel et~al.(2012)Kinzel, Mulligan, and Dabiri}]{Kinzel2012}
\bibinfo{author}{M.~Kinzel}, \bibinfo{author}{Q.~Mulligan},
  \bibinfo{author}{J.~O. Dabiri},
\newblock \bibinfo{title}{{Energy exchange in an array of vertical-axis wind
  turbines}},
\newblock \bibinfo{journal}{Journal of Turbulence} \bibinfo{volume}{14}
  (\bibinfo{year}{2012}) \bibinfo{pages}{N38}.
\bibitem[{Barthelmie et~al.(2007)Barthelmie, Frandsen, Nielsen, Pryor, Rethore,
  and Jørgensen}]{Barthelmie2007}
\bibinfo{author}{R.~J. Barthelmie}, \bibinfo{author}{S.~T. Frandsen},
  \bibinfo{author}{M.~N. Nielsen}, \bibinfo{author}{S.~C. Pryor},
  \bibinfo{author}{P.~E. Rethore}, \bibinfo{author}{H.~E. Jørgensen},
\newblock \bibinfo{title}{{Modelling and measurements of power losses and
  turbulence intensity in wind turbine wakes at Middelgrunden offshore wind
  farm}},
\newblock \bibinfo{journal}{Wind Energy} \bibinfo{volume}{10}
  (\bibinfo{year}{2007}) \bibinfo{pages}{517--528}.

\end{thebibliography}
\end{document}